\documentclass[11pt]{article}
\usepackage{amsmath,amsthm,amssymb,amsfonts,cite,enumerate}
\usepackage[a4paper,left=0.5in,right=1in]{geometry}
\usepackage[usenames]{color}
\usepackage{graphicx}
\usepackage{epsfig}
\usepackage{dcolumn}
\usepackage{bm}
\usepackage{authblk} 
\usepackage{soul} 
\usepackage{array,multirow} 

\usepackage{pifont} 

\begin{document}

\def \d {{\rm d}}

\def \bm #1 {\mbox{\boldmath{$m_{(#1)}$}}}

\def \bF {\mbox{\boldmath{$F$}}}
\def \bV {\mbox{\boldmath{$V$}}}
\def \bff {\mbox{\boldmath{$f$}}}
\def \bT {\mbox{\boldmath{$T$}}}
\def \bk {\mbox{\boldmath{$k$}}}
\def \bl {\mbox{\boldmath{$\ell$}}}
\def \bn {\mbox{\boldmath{$n$}}}
\def \bbm {\mbox{\boldmath{$m$}}}
\def \tbbm {\mbox{\boldmath{$\bar m$}}}
\def \xib {\mbox{\boldmath{$\xi$}}}

\def \bo {\mbox{\boldmath{$\omega$}}}

\def \T {\bigtriangleup}
\newcommand{\msub}[2]{m^{(#1)}_{#2}}
\newcommand{\msup}[2]{m_{(#1)}^{#2}}

\newcommand{\be}{\begin{equation}}
\newcommand{\ee}{\end{equation}}

\newcommand{\beq}{\begin{eqnarray}}
\newcommand{\eeq}{\end{eqnarray}}
\newcommand{\pa}{\partial}
\newcommand{\pp}{{\it pp\,}-}
\newcommand{\ba}{\begin{array}}
\newcommand{\ea}{\end{array}}

\newcommand{\M}[3] {{\stackrel{#1}{M}}_{{#2}{#3}}}
\newcommand{\m}[3] {{\stackrel{\hspace{.3cm}#1}{m}}_{\!{#2}{#3}}\,}

\newcommand{\tr}{\textcolor{red}}
\newcommand{\tb}{\textcolor{blue}}
\newcommand{\tg}{\textcolor{green}}

\newcommand*\bg{\ensuremath{\boldsymbol{g}}}
\newcommand*\bE{\ensuremath{\boldsymbol{E}}}
\newcommand*\bh{\ensuremath{\boldsymbol{h}}}

\def\a{\alpha}
\def\g{\gamma}
\def\de{\delta}

\def\b{\beta}

\def\E{{\cal E}}
\def\B{{\cal B}}
\def\R{{\cal R}}
\def\F{{\cal F}}
\def\L{{\cal L}}

\def\e{e}
\def\bb{b}

\newtheorem{theorem}{Theorem}[section] 
\newtheorem{cor}[theorem]{Corollary} 
\newtheorem{lemma}[theorem]{Lemma} 
\newtheorem{prop}[theorem]{Proposition}
\newtheorem{definition}[theorem]{Definition}
\newtheorem{remark}[theorem]{Remark}

\title{New look at AdS black holes with conformal scalar hair}

\author[1]{Marcello Ortaggio\thanks{ortaggio(at)math(dot)cas(dot)cz}}

\affil[1]{Institute of Mathematics, Czech Academy of Sciences, \newline \v Zitn\' a 25, 115 67 Prague 1, Czech Republic}

\maketitle

\begin{abstract}
 We revisit static, spherically symmetric solutions to AdS-Einstein gravity with a conformally coupled scalar field (and no self-interaction potential) in four dimensions. We first observe that a convenient choice of coordinates leads to a significant simplification of the field equations, which enables one to identify various roots of the indicial equations and thus distinct branches of solutions. Next, we construct an explicit 2-parameter hairy black hole solution in terms of an infinite power series around the event horizon. The black hole is non-extremal with a regular scalar field on and outside the event horizon, and it reduces to the Schwarzschild-AdS metric in the limit of vanishing hair. Its properties are illustrated for various values of the parameters and compared with previous numerical results by other authors. In addition, the analysis reveals the presence of a photon sphere and how  the scalar field affects its size and the angular radius of the corresponding shadow. The thermodynamics of the solution is also briefly discussed.
\end{abstract}


\section{Introduction}

\label{sec_intro}

According to the well-known ``no-hair conjecture'' \cite{RufWhe71}, the endpoint of gravitational collapse is expected to be a black hole completely characterized by its conserved charges -- any other independent parameter (``hair'') cannot be supported. While this is sustained by various no-hair theorems in electrovac General Relativity, numerous hairy black holes have been constructed in more general contexts (cf., e.g., the reviews \cite{Heuslerbook,ChrCosHey12} and \cite{Bekenstein96,VolGal99,HerRad15,Volkov17} on uniqueness theorems and hairy black holes, respectively).

In particular, no-hair theorems can be bypassed if one considers a scalar field conformally coupled to gravity. Among other reasons, the interest in conformal scalar fields stems from the fact that their stress-energy has interesting properties from a quantum viewpoint \cite{CalColJac70}, and it has been considered as a source in semi-classical General Relativity (e.g. in \cite{HuPar78}) and in inflationary models (cf., e.g., \cite{Faraonibook} for references). An early (static, spherically symmetric, asymptotically flat) black hole supporting a conformal scalar was discovered by Bocharova, Bronnikov, Melnikov and Bekenstein (BBMB) \cite{BocBroMel70,Bekenstein74}. Within the same theory, an exact spherical solution in the presence of a cosmological constant $\Lambda>0$ has been obtained more recently in \cite{MarTroZan03} (dubbed MTZ in what follows), and subsequently extended to topological black holes with $\Lambda<0$ in \cite{MarTroZan04,MarTroSta06}. However, the solutions   
 \cite{MarTroZan03,MarTroZan04,MarTroSta06} require admitting also a fine-tuned quartic self-interaction (or a constant scalar field), and do not possess an independent hair parameter. It would thus be desirable to clarify if black holes with conformal hair and non-zero $\Lambda$ exist also when no self-interaction is included in the theory. In such a case, according to a no-hair theorem of \cite{Winstanley03}, a black hole solution conformally coupled to a scalar field with $\Lambda>0$ cannot be asymptotically de~Sitter. However, the existence of static, spherical black holes with $\Lambda<0$ and without self-interaction has been demonstrated numerically in \cite{Winstanley03} (see also \cite{RadWin05}).\footnote{Refs.~\cite{Winstanley03,RadWin05} contain also results in the presence of a mass term or a self-interaction potential for the scalar field, as well as topological and higher-dimensional black holes. Some of these solutions have also been shown to be linearly stable under spherically symmetric perturbations \cite{Winstanley03,RadWin05}.} But constructing a corresponding exact solution seems a challenging task, also because one needs two independent metric functions\footnote{See \cite{PodOrt06,PodSva15,HerOrt20} for the invariant meaning of this fact in terms of null alignment properties of the curvature, and \cite{Jacobson07} for related observations.} to characterize such black holes \cite{Winstanley03,RadWin05} (as opposed to the solutions of \cite{BocBroMel70,Bekenstein74,MarTroZan03,MarTroZan04,MarTroSta06}). It is the purpose of the present contribution to reconsider this problem analytically, thus providing analytic support to the numerical findings of \cite{Winstanley03,RadWin05}, as well as adding several new observations (in particular regarding the presence of a photon sphere and the thermodynamics).

The plan of the paper is as follows. In the remaining part of this section we present the theory under consideration and the corresponding field equation. In section~\ref{sec_ansatz}, taking advantage of conformal properties of the theory and following \cite{Ray24}, we first cast the reduced field equations for a static, spherically symmetric ansatz in a convenient form that is considerably simpler than the one obtained in the standard Schwarzschild-like coordinates. Those will be therefore amenable to an analytic treatment, and we describe how to solve them in terms of infinite power series. This will reveal the existence of distinct branches of solutions characterized by various roots of the indicial equations and different numbers of free parameters (integration constants). By focusing on an expansion around an event horizon, in section~\ref{sec_010} we construct an explicit black hole solution characterized by 2 parameters (horizon radius and scalar hair), in addition to a negative $\Lambda$. Its mathematical (convergence) and physical properties (vacuum limit, asymptotic behavior, photon sphere and shadow) are discussed and further illustrated by several graphs. Some comments on the thermodynamics are also provided. Concluding remarks are given in section~\ref{sec_concl}. Appendix~\ref{app_energy} briefly discusses the energy-momentum tensor and some of the energy conditions for the hairy black hole solution of section~\ref{sec_010}.

We will consider the theory
\be
 I=\frac{1}{16\pi}\int\d^4 x\sqrt{-g}\left(R-2\Lambda-S\right) ,
 \label{I}
\ee
where $S$ is the full trace of the following Riemann-like tensor \cite{OliRay12}
\be
 S^{cd}_{ab}=\phi^2R^{cd}_{ab}+\delta^{[c}_{[a}\left[-4\phi\nabla^{d]}\nabla_{b]}\phi+8(\nabla_{b]}\phi)\nabla^{d]}\phi-2\delta^{d]}_{b]}(\nabla_{e}\phi)\nabla^{e}\phi   \right] ,
\ee
i.e.,\footnote{The contribution to the integral~\eqref{I} from the term $6\phi\Box\phi$ can be also written, up to a boundary term, as a standard kinetic term $-6(\nabla_a\phi)\nabla^a\phi$, as in, e.g., \cite{CheTag68,CalColJac70,BocBroMel70,Deser70,Parker73}.\label{footn_kinet}}
\be
 S=S^{ab}_{ab}=\phi^2R-6\phi\Box\phi .
 \label{S}
\ee

Noticing that $\tilde R^{cd}_{ab}=\phi^{-4}S^{cd}_{ab}$ corresponds to the Riemann tensor of a conformally rescaled metric
\be
 \tilde g_{ab}=\phi^2g_{ab} ,
\ee
and $\tilde R^{c}_{a}=\phi^{-4}S^{c}_{a}=\phi^{-4}S^{cb}_{ab}$ to its Ricci tensor (cf., e.g., \cite{Waldbook}), the variations of the action~\eqref{I} giving rise to the tensorial and scalar field equations $E_{ab}=0$, $E_{\mbox{s}}=0$ can be written compactly as
\beq
  & & E_{ab}=G_{ab}+\Lambda g_{ab}-\phi^2\tilde G_{ab} , \label{Einstein} \\
	& & E_{\mbox{s}}=-2\phi^3\tilde R , \label{scalar} 
\eeq
where $G_{ab}=R_{ab}-\frac{R}{2}g_{ab}$ and $\tilde G_{ab}=\tilde R_{ab}-\frac{\tilde R}{2}\tilde g_{ab}=\phi^{-2}(S_{ab}-\frac{S}{2}g_{ab})$ are the Einstein tensors of $g_{ab}$ and $\tilde g_{ab}$, respectively. The term $\tilde G_{ab}$ in~\eqref{Einstein} is traceless after imposing $E_{\mbox{s}}=0$ with~\eqref{scalar}, so that the trace of~\eqref{Einstein} gives $-E^a_{\phantom{a}a}=R-4\Lambda$  (and thus $R=4\Lambda$ on shell).

\section{Ansatz and reduced field equations}

\label{sec_ansatz}

\subsection{Conformally Kundt coordinates and field equations}

\label{subsec_confK}

In recent works on quadratic gravity \cite{PraPraPodSva17,Podolskyetal18}, it has been pointed out that a convenient choice of coordinates can lead to a drastic simplification of the field equations of that theory, which proved extremely useful in the construction of black hole solutions. Since the quadratic gravity action contains a conformally invariant term, one might expect that the coordinates of \cite{PraPraPodSva17,Podolskyetal18} could be suitable also in other theories possessing some kind of (at least partial) conformal invariance. As observed in \cite{Ray24}, this turns out to be the case, in particular, for the theory~\eqref{I}, and in the following we thus employ the conformal Kundt coordinates defined in \cite{PraPraPodSva17,Podolskyetal18}. 

We consider a metric of the form
\be
 \d s^2=\Omega^2\left(-2\d u\d r+H\d u^2+\d\Sigma^2\right) , \qquad \d\Sigma^2=2P^{-2}\d\zeta\d\bar\zeta , \qquad P=1+\frac{1}{2}\zeta\bar\zeta ,
 \label{metric}
\ee
where $\d\Sigma^2$ is a 2-sphere of constant Gaussian curvature normalized to 1, and we assume that $\Omega$, $H$ and the scalar field $\phi$ depend only on $r$. This ansatz thus includes all static metrics with spherical symmetry. The Killing vector field $\pa_u$ is timelike where $H<0$. In the following we shall exclude from the analysis the special case $\Omega$=const, which does not describe black holes since it corresponds to Kundt metrics \cite{Stephanibook}. 
Notice that a coordinate transformation \cite{Stephanibook,Podolskyetal18}
\be
 u=\alpha^{-1}\hat u , \qquad r=\alpha \hat r+\beta ,
 \label{gauge}
\ee
leaves the metric~\eqref{metric} unchanged if $H$ is also rescaled as $H=\alpha^2\hat H$ ($\alpha\neq0$ and $\beta$ are constants). 

We further observe that $\Omega^{-1}=0$ defines conformal infinity, provided $\nabla\Omega^{-1}\neq0$ (and $H$ is regular) there, and the considered spacetimes are thus asymptotically simple (at least locally) \cite{penrosebook2}.\footnote{Strictly speaking, this is true if, within the spacetime, one can get arbitrarily close to the hypersurface $\Omega^{-1}=0$ (without, e.g., encountering a singularity). The specific character of conformal infinity is determined by the asymptotic sign of $H$ (i.e., Minkowskian if $H\to0$, and (A)dS if $H$ tends to a (negative) positive value).}
  Note also that the Weyl invariant \cite{Podolskyetal18}
\be
 C_{abcd}C^{abcd}=\frac{1}{3}(2+H'')^2\Omega^{-4} ,
 \label{C2}
\ee
signals the presence of a possible curvature singularity at $\Omega=0$. The spacetime is conformally flat iff $2+H''=0$, and of constant curvature if, additionally, $(\Omega^{-1})''=0$ (hereafter a prime denotes differentiation w.r.t. $r$). 

It is also interesting to point out that the coordinates~\eqref{metric} appear to be natural for the study of {\em photon spheres} \cite{ClaVirEll01,VirEll02}, which are located at $r=r_{ps}$ such that $H'(r_{ps})=0$ with $H(r_{ps})<0$.\footnote{The definition of a photon spheres for static, spherically symmetric spacetimes in Schwarzschild coordinates~\eqref{metric_Schw}, \eqref{functions_Schw} reads $2f=\rho f_{,\rho}$ with $g>0$, where both conditions must hold at the photon surface radius $\rho=\rho_{ps}$, see \cite{Atkinson65} (cf. also \cite{ClaVirEll01,HasPer02,VirEll02,Bozza02}). To our knowledge, the use of the conformal Kundt coordinates~\eqref{metric} in this context had not been considered in the literature so far -- however, a ``potential'' which equals (up to a sign) our metric function $H$ was conveniently defined in \cite{HasPer02}.\label{footn_ps}} If a static observer sits at $r=r_O$ in the (static) exterior region, the angular radius $\chi_{_O}$ of the {\em black hole shadow} is given by\footnote{This follows readily from eq.~(43,\cite{PerTsuBis15}) keeping into account the comments in footnote~\ref{footn_ps}, but an equivalent formula was given earlier in \cite{PanDur86} (cf. also~\cite{CveGibPop16}). For simplicity, here we have assumed there is a single photon sphere in the exterior region (as will be indeed the case in the rest of the paper, cf. section~\ref{sec_010}). A more general discussion, including the angular radius of shadows defined for comoving observers in asymptotically de~Sitter spacetimes, can be found in  \cite{PerTsuBis18} -- cf. also the review \cite{PerTsu22} and references therein. In the special case of the Schwarzschild black hole, formula~\eqref{shadow} was first obtained in \cite{ZelNov65,Synge66}, and extended to include $\Lambda$ in \cite{StuHle99} (see again \cite{PerTsu22} for more references and for an overview of the various terminology used in this context in the literature).}
	\be
	 \sin^2\chi_{_O}=\frac{H(r_O)}{H(r_{ps})} .
	 \label{shadow}
	\ee
These observations will be useful in section~\ref{subsec_prop}.

As in \cite{Ray24}, it proves convenient to introduce a rescaled scalar field 
\be
 \Psi=\Omega\phi .
 \label{Psi}
\ee
Using the latter, for the ansatz~\eqref{metric} the non-zero components of the field equations~\eqref{Einstein}, \eqref{scalar} read
\beq
& & \frac{1}{2}\Omega^6E^{uu}=\Omega^3(\Omega^{-1})''-\Psi^3(\Psi^{-1})'' , \label{uu} \\
& & 2P^{-2}\Omega^6E^{\zeta\bar\zeta}=\left[H(\Omega^2-\Psi^2)'\right]'+2\Lambda\Omega^4-H''(\Omega^2-\Psi^2)-6\Omega(H\Omega')'+6\Psi(H\Psi')'   ,
\label{Zz} \\
& & \Omega^6E^{ur}=(\Omega^2-\Psi^2)-\Lambda\Omega^4-(\Omega^{-1})'(\Omega^3H)'+(\Psi^{-1})'(\Psi^3H)'  ,   \label{ur} \\
& & -\frac{1}{2}\Omega^3E_{\mbox{s}}=6(H\Psi')'+\Psi(2+H'') , \label{scalar3} 
\eeq
while $E^{rr}$ is proportional to $E^{ur}$.

We observe that once~\eqref{scalar3} is satisfied the conservation equation \cite{HorLov72} gives $E^{ab}_{\phantom{ab};b}=0$, which implies that \eqref{Zz} is not an independent equation (except in the excluded Kundt case $\Omega=$const). Keeping this into account and replacing~\eqref{ur} by the linear combinations $E^{ur}-\frac{1}{2}HE^{uu}$, the system of independent equations to be solved can thus be written in the simplified form 
\beq
 & & \Omega^3(\Omega^{-1})''=\Psi^3(\Psi^{-1})'' , \label{uu_2} \\
 & & (\Omega^2-\Psi^2)+\textstyle{\frac{1}{2}}\left[H(\Omega^2-\Psi^2)'\right]'-\Lambda\Omega^4=0  ,   \label{ur_comb} \\
 & & 6(H\Psi')'+\Psi(2+H'')=0 . \label{scalar4} 
\eeq

This is more compact than the corresponding system in Schwarzschild coordinates \cite{Winstanley03,RadWin05,DotGleMar08} (defined in section~\ref{subsec_Schwar} below), and as a further advantage it is an autonomous system (see \cite{Podolskyetal18} for similar comments in a different context).

Let us further note that two other (non-independent) simple equations can be obtained which will be useful for practical purposes in the following. First, the trace of the Einstein equation reads 
\be
\Omega^4E^a_{\phantom{a}a}=4\Lambda\Omega^4-(\Omega^2-\Psi^2)(2+H'')-6\Omega(H\Omega')'+6\Psi(H\Psi')' ,
\ee	
so that imposing $E^a_{\phantom{a}a}=0$ with~\eqref{scalar4} gives
\be
6(H\Omega')'+\Omega(2+H'')-4\Lambda\Omega^3=0 . \label{eqR}
\ee	
Apart from the $\Lambda$ term, this is a counterpart of~\eqref{scalar4}, upon interchanging $\Omega\leftrightarrow\Psi$ (in the $\Lambda=0$ case, this gives rise to the ``duality'' first observed in \cite{Bekenstein74} and discussed recently in arbitrary dimensions in \cite{MarNoz21,Ray24}).

In addition, using~\eqref{scalar4}  and \eqref{eqR}, the linear combination $E^{ur}-HE^{uu}=0$ can be written as 
\be
[H(\Omega^2-\Psi^2)]''-2\Lambda\Omega^4=0 .
\ee

The latter equation immediately reveals that for $H=0$ one obtains $\Lambda=0$, therefore throughout the paper we can assume $H\neq0$, since we are interested in solutions with $\Lambda\neq0$. For the same reason we can assume $\Omega^2\neq\psi^2$.

\subsection{Schwarzschild coordinates}

\label{subsec_Schwar}

For later discussion, let us note that metric~\eqref{metric} can be cast in standard Schwarzschild-like coordinates 
\be
 \d s^2=-f(\rho)\d t^2+\frac{d\rho^2}{g(\rho)}+\rho^2\d\Sigma^2 ,
 \label{metric_Schw}
\ee
via the transformation \cite{PraPraPodSva17} 
\be
 \d t=-\d u+\frac{\d r}{H}, \qquad \rho=\Omega ,  
 \label{transf_Schw}
\ee
giving 
\be
 f=-\Omega^2 H  , \qquad g=-\left(\frac{\Omega'}{\Omega}\right)^2 H.
 \label{functions_Schw}
\ee
In regions where $H<0$, $\pa_t$ is timelike and throughout the paper it will be assumed to be future oriented (so that $\pa_u=-\pa_t$ is past oriented).

We observe that the above coordinate transformation is not defined if $\Omega$=const (corresponding to Kundt metrics) or $H=0$ -- those cases are, however, not relevant to the present paper, as mentioned above.

\subsection{Power series expansions and summary of solutions}

Similarly as in \cite{Podolskyetal18}, in order to construct analytic solutions, we will employ a Frobenius-like method (cf., e.g., \cite{Piaggiobook}), which consists in expanding the unknown functions around an arbitrary, finite value of the radial coordinate $r=r_0$ as infinite power series
\be
 \Omega=\Delta^n\sum_{k=0}a_k\Delta^k , \qquad \Psi=\Delta^q\sum_{k=0}b_k\Delta^k , \qquad H=\Delta^p\sum_{k=0}c_k\Delta^k ,
 \label{expansions}
\ee
where $\Delta\equiv r-r_0$, and $(n,p,q)$ and $a_k$, $b_k$ and $c_k$ are constants, with $a_0,b_0,c_0\neq0$. Then, by plugging~\eqref{expansions} into the field equations~\eqref{uu_2}--\eqref{scalar4}, one is first typically able to constraint the permitted values of the exponents $(n,p,q)$ by considering the terms of lowest orders in $\Delta$ in the resulting equations. The next step consists in using higher orders to construct a set of recurrence formulas such that the coefficients $a_k$, $b_k$ and $c_k$ in~\eqref{scalar4} for an arbitrary $k$ are determined in terms of those with lower indices. This enables one to identify the coefficients which remain arbitrary as integration constants (the details depending on the specific solution under consideration), and to evaluate the metric functions as accurately as desired at any point within the convergence radius of the series.

Different values of $(n,p,q)$ (i.e., the ``indices''\cite{Piaggiobook}) correspond to different branches of solutions. The derivation of all possible cases is straightforward but lengthy and we shall present a thorough analysis elsewhere. For the purposes of the present contribution, it suffices to only summarize here the resulting possibilities in table~\ref{tab_exponents}. Among those, it is easy to identify the branch $(n=0,p=1,q=0)$ as an expansion around a non-extremal Killing horizon -- these are thus solutions which may describe black holes and will be analyzed in the rest of the paper. We also observe that an expansion around a degenerate horizon would require $n=0$ and $p=2$ -- no such solutions are thus present in the class constructed with the above method.\footnote{This is due to the assumption $\Lambda\neq0$ -- in the case $\Lambda=0$ one would recover the well-known extremal solution of \cite{BocBroMel70,Bekenstein74}.} The remaining cases will be analyzed elsewhere.

\begin{table}[htbp]
	\[
	\begin{array}{|c||c|c|c|} \hline
		\  (n,p,q)\  & \mbox{ expansion at  } & \mbox{ behaviour of $\phi$ }  & \mbox{ parameters } \\ \hline\hline
		(-1,0,0) & \mbox{ conformal infinity ($\rho\to\infty$) } & \phi\sim\rho^{-1} &  3 \\ \hline      
		(-1,0,1)  & \mbox{ conformal infinity ($\rho\to\infty$) } & \phi\sim\rho^{-2} & 2 \\ \hline      
		(0,0,0)  &    \mbox{ generic point } & \phi\to\mbox{const } & 4 \\ \hline
		(0,0,1)  &  \mbox{ zero of } \phi & \phi\to0 & 3 \\ \hline      
		(0,1,0) &  \mbox{ non-extremal horizon } & \mbox{ $\phi\to$const } & 2 \\ \hline
		(1,0,0)  &  \mbox{ non-Schwarzschildian singularity } & \phi\to\infty  & 3 \\ \hline
	\end{array}
	\]
	\caption{\small~Summary of possible values of the exponents $(n,p,q)$ for solutions with $\Lambda\neq0\neq\phi$, ordered by increasing value of $n$. The last column indicates the number of essential (i.e., after using the gauge freedom~\eqref{gauge}) integration constants which characterize a given branch of solutions. It should be emphasized that, for a particular (infinite) set of $n\in\mathbb{N}^+$, there exists also a special branch with $n=q>0$, $p<2$ and $a_0^2-b_0^2=0$ which has been omitted from this table. Such solutions do not represent expansions near a black hole horizon and will be discussed elsewhere.}
	\label{tab_exponents}
\end{table}

Let us further notice that the scaling freedom~\eqref{gauge} rescales the coefficients in~\eqref{expansions} as $\hat a_k=\alpha^{n+k}a_k$, $\hat b_k=\alpha^{q+k}b_k$ and $\hat c_k=\alpha^{p-2+k}c_k$, and thus always allows one to normalize arbitrarily one of those (provided it is non-vanishing), along with $\hat r_0=\alpha^{-1}(r_0-\beta)$. This will be useful in the following to get rid of one unphysical parameter and write the solutions in a canonical form.

\section{Black hole solutions: case $(n=0,p=1,q=0)$}

\label{sec_010}

The focus of this paper is on black holes with a non-extremal horizon. From now on we thus study the branch of solutions with exponents $(n=0,p=1,q=0)$, which means we are expanding the metric near a non-extremal Killing horizon, located at $r=r_0$. The particular value $r_0$ can be shifted at will by choosing $\beta$ in~\eqref{gauge} arbitrarily and has therefore no physical meaning. The physical (dimensionful) horizon radius is given in Schwarzschild coordinates~\eqref{metric_Schw} by the integration constant $a_0$ (cf. more comments in the following) 
\be
	\rho(r_0)=a_0>0 .
 \label{a_0}
\ee 
The sign of $a_0$ has been fixed (without losing generality) thanks to the invariance of~\eqref{metric} under $\Omega\to-\Omega$.

By taking an appropriate sign of $\alpha$ in~\eqref{gauge} one can further also set $a_1>0$, such that near and across the horizon $\rho$ is monotonically increasing with $r$.\footnote{Except in the special case $a_1=0$, which we will not consider in the following.} Since our aim is to study the spacetime in the vicinity of an outer black hole horizon, we then need to assume $c_0<0$ (which ensures that $\pa_u$ is timelike, cf.~\cite{PraPraOrt23} for related comments). Hereafter we therefore restrict ourselves to the parameter range
\be
	a_1>0 , \qquad c_0<0 .
	\label{hor_conds}
\ee

Let us emphasize that the solution obtained in the following admits any sign of $\Lambda$ (see section~\ref{subsec_stealth_Schw} for $\Lambda=0$). However, because of the result of \cite{Winstanley03} mentioned in section~\ref{sec_intro}, in the discussion of black holes we will restrict ourselves to the case $\Lambda<0$.

\subsection{General solution with $\Lambda\neq0$: hairy AdS black hole}

\label{subsec_010_general}

Substituting~\eqref{expansions} with $(n=0,p=1,q=0)$ into~\eqref{uu_2}--\eqref{scalar4} reveals that a solution in this branch is possible only for $a_0^2-b_0^2\neq0$, which will thus be assumed hereafter. According to the value $b_0$ of the scalar field $\Psi$ at the horizon, one can thus identify two disconnected branches of solutions, i.e., those with $a_0^2-b_0^2>0$ or with $a_0^2-b_0^2<0$. In the following we will be focusing mostly on the former, since it is the only one continuously connected (for $b_0\to0$) to the vacuum AdS black holes (and the only one giving rise a positive entropy, as discussed later on in section~\ref{subsubsec_thermod}).

At the lowest orders \eqref{uu_2}--\eqref{scalar4} further give
\be
 a_1=-\frac{a_0}{c_0}+\frac{a_0^3 \Lambda  \left(3 a_0^2-2 b_0^2\right)}{3 c_0 \left(a_0^2-b_0^2\right)} , \qquad a_2=\frac{a_0}{c_0^2}-\frac{2 a_0^3 \Lambda  \left(3 a_0^2-2 b_0^2\right)}{3 c_0^2 \left(a_0^2-b_0^2\right)}+\frac{a_0^5 \Lambda ^2 \left(3 a_0^2-b_0^2\right)}{3 c_0^2 \left(a_0^2-b_0^2\right)} , 
 \label{a1}
\ee
\be
	c_1=2-\frac{a_0^4 \Lambda }{a_0^2-b_0^2} , \qquad c_2=\frac{1}{c_0}-\frac{4 a_0^4  \Lambda }{3 c_0 \left(a_0^2-b_0^2\right)}+\frac{a_0^6 \Lambda^2 \left(3 a_0^2+4 b_0^2\right)}{9 c_0 \left(a_0^2-b_0^2\right)^2} ,
\ee
\be
	b_1=-\frac{b_0 }{c_0}+\frac{a_0^4 b_0 \Lambda }{3 c_0 \left(a_0^2-b_0^2\right)} , \qquad b_2=\frac{b_0 }{c_0^2}+\frac{a_0^4 b_0 \Lambda  \left(a_0^2 \Lambda -6 \right)}{9 c_0^2 \left(a_0^2-b_0^2\right)} .
\ee

All higher order coefficients are then obtained recursively as
\be
	k (k-1) \left(a_0 a_k-b_0 b_k\right)=\sum _{j=1}^{k-1} j (2 k-3 j+1) \left(a_j a_{k-j}-b_j b_{k-j}\right) ,
\label{a_k}	
\ee
\be
 k(k+1)b_0 c_k=-6 k^2c_0b_k-2 b_{k-1}-\sum _{j=1}^{k-1} c_j \left[6 k (k-j)+j(j+1) \right] b_{k-j} ,
 \label{c_k}
\ee
\beq
 6k^2c_0 \left(a_0^2-b_0^2\right) b_k=2a_0 \left(b_0 a_{k-1}-a_0 b_{k-1}\right)-4a_0 b_0 \Lambda  \sum _{j=0}^{k-1}\sum _{l=0}^{k-j-1} a_j a_l a_{k-j-l-1} \nonumber\\ 
	{}+\sum _{j=1}^{k-1}\Bigg [\frac{6k}{k-1}j (2k-3 j+1) \left(a_j a_{k-j}-b_j b_{k-j}\right)b_0 c_0 \label{b_k} \nonumber \\
	{}+\left[6 k (k-j)+j (j+1)\right] a_0c_j \left(b_0 a_{k-j}-a_0 b_{k-j}\right)\Bigg]  ,
\eeq
which thus fully determine (via~\eqref{expansions}) the solution.

The three coefficients $a_0$, $b_0$ and $c_0$ remain arbitrary, but the modulus of $c_0$ can be rescaled as desired using~\eqref{gauge} with $\alpha>0$ (recall~\eqref{hor_conds}). There eventually remains two independent free physical parameters related to mass and scalar hair. Because of~\eqref{hor_conds}, they must obey
\be
  1-\Lambda a_0^2-\frac{\Lambda}{3}\frac{a_0^2b_0^2}{a_0^2-b_0^2}>0 .
 	\label{hor_conds2}
\ee

For definiteness, it may also be useful to give the leading terms of the above solution in the Schwarzschild coordinates~\eqref{metric_Schw}. This gives rise to an expansion at the horizon radius $\rho=a_0$, which is given, up to the subleading order, by 
\beq
 & & f=\frac{3 a_0 c_0^2 \left(a_0^2-b_0^2\right)}{3 \left(a_0^2-b_0^2\right)-a_0^2\Lambda  \left(3 a_0^2-2  b_0^2\right)}(\rho-a_0) \nonumber \\
 & &	\qquad {}-3 c_0^2 \left(a_0^2-b_0^2\right) \frac{a_0^4 b_0^2 \Lambda ^2 \left(6 a_0^2-5 b_0^2\right)+9 \left(a_0^2-b_0^2\right) \left[ \left(a_0^2-b_0^2\right)-a_0^4 \Lambda \right]}{\left[3 \left(a_0^2-b_0^2\right)-a_0^2\Lambda  \left(3 a_0^2-2  b_0^2\right)\right]^3}(\rho-a_0)^2+\ldots , \nonumber \\
 & & g=\left[\frac{1}{a_0}-\frac{\Lambda}{3}\frac{a_0 \left(3 a_0^2-2 b_0^2\right)}{a_0^2-b_0^2}\right](\rho-a_0) \\
 & &	\qquad {}-\frac{a_0^4 b_0^2 \Lambda ^2\left(6 a_0^2-b_0^2\right)+9 \left(a_0^2-b_0^2\right) \left[\left(a_0^2-b_0^2\right)-a_0^4 \Lambda \right]}{3 a_0^2 \left(a_0^2-b_0^2\right) \left[3 \left(a_0^2-b_0^2\right)-a_0^2 \Lambda  \left(3 a_0^2-2 b_0^2\right)\right]}(\rho-a_0)^2+\ldots , \nonumber \\
 & & \phi=\frac{b_0}{a_0}+\frac{2 b_0 \Lambda \left(a_0^2-b_0^2\right)}{3 \left(a_0^2-b_0^2\right)-a_0^2 \Lambda  \left(3 a_0^2-2 b_0^2\right)}(\rho-a_0)+\ldots , \nonumber 
\eeq
where \eqref{hor_conds2} ensures that the above denominators are non-zero. At the leading order, $f$ and $g$ coincide if one chooses the gauge $a_1=a_0^2$ (recall~\eqref{a1}). In the special case $1-\frac{\Lambda}{3}a_0^2(3a_0^2-2b_0^2)(a_0^2-b_0^2)^{-1}=0$ (i.e., $a_1=0$), expansions in Schwarzschild coordinates need to be treated differently due to the appearance of non-integers powers.

As mentioned in section~\ref{sec_intro}, a result of \cite{Winstanley03} implies that, for $\Lambda>0$, the above black holes are not asymptotically de~Sitter. However, numerics in the case $\Lambda<0$ shows that they are compatible with anti-de~Sitter asymptotics \cite{Winstanley03,RadWin05}. Before further elucidating their properties, in the following two subsections we briefly comment on the solutions of \eqref{Einstein}, \eqref{scalar} (of the form \eqref{metric}, \eqref{expansions}) possessing an horizon (also) in the limits $\Lambda=0$ and $\phi=0$.

\subsubsection{Limit $\Lambda=0$: Schwarzschild (stealth) solution}

\label{subsec_stealth_Schw}

By setting $\Lambda=0$, one can easily prove by induction that the coefficients $a_k$, $b_k$ and $c_k$ obtained in section~\ref{subsec_010_general} satisfy 
\beq
& & a_k=a_0\left(-\frac{1}{c_0}\right)^k , \qquad b_k=b_0\left(-\frac{1}{c_0}\right)^k \qquad (k\ge0) , \nonumber \\
& & c_1=2 , \qquad c_2=\frac{1}{c_0} ,  \qquad c_l=0 \qquad (l\ge3) . \label{Schw_limit}
\eeq
This gives rise to series that can be summed up exactly to obtain the Schwarzschild solution with $2M=a_0$ and a constant scalar field $\phi=b_0/a_0$ (after using~\eqref{gauge} to set $r_0=-1/a_0$, $c_0=-1/a_0$, see also section~VIII.B.1 of \cite{PodSvaPraPra20} for a related discussion). This is a stealth solution since $T_{ab}=0$. The outer horizon condition~\eqref{hor_conds2} is automatically satisfied.

This is in agreement with the known result \cite{XanZan91,XanDia92,Klimcik93,Zannias95} that there exist no non-extremal (non-stealth) black holes in the case $\Lambda=0$.

\subsubsection{Limit $\phi=0$: vacuum Schwarzschild-(A)dS black hole}

\label{subsec_SchwAdS}

In the derivation of section~\ref{subsec_010_general} we assumed $\psi\neq0$. For a comparison, let us observe that the vacuum (A)dS black hole can in a similar way be obtained as a series expansion by solving the equations~\eqref{uu_2}--\eqref{scalar4} with $\psi=0$ (as discussed in \cite{PraPraPodSva21}). This gives $a_k=a_0\left(a_1/a_0\right)^k$ with $a_1=a_0(\Lambda a_0^2-1)/c_0$, and $c_1=2-\Lambda a_0^2$, $c_2=(\Lambda a_0^2-3)(\Lambda a_0^2-1)/(3c_0)$, while $c_k=0$ for $k\ge3$.\footnote{Here we assume $\Lambda a_0^2-1\neq0$ to avoid solutions of the Kundt class (for which $\Omega=$const). The special case $\Lambda a_0^2-1=0$ gives rise to the (anti-)Nariai spacetime (cf.~\cite{PraPraOrt23}).} The parameters $r_0$, $a_0$ and $c_0$ remain arbitrary, but one can use \eqref{gauge} to set $r_0=-a_0^{-1}$ and $c_0=(\Lambda a_0^2-1)/a_0$ (i.e., $a_1=a_0^2$), thereby obtaining the standard normalization $\Omega=-1/r$, $H=-r^2+\Lambda/3+\frac{a_0}{3}(\Lambda a_0^2-3)r^3$, where the usual mass parameter is thus given by $2M=a_0(1-\frac{\Lambda}{3}a_0^2)$. For $M>0$, a photon sphere is located at $r^{-1}=-3M$, i.e., at $\rho=3M$ \cite{StuHle99,ClaVirEll01} in the coordinates~\eqref{metric_Schw}.

Here~\eqref{hor_conds2} gives
\be
1-\Lambda a_0^2>0 . 
\label{hor_conds_Einst}
\ee 
Saturating the inequality in ~\eqref{hor_conds_Einst} (i.e., taking the limit $1-\Lambda a_{0}^2\to0$) gives rise to the standard extremality condition for dS black holes \cite{GibHaw77}. When $1-\Lambda a_0^2<0$, condition~\eqref{hor_conds2} is violated, and the horizon $r=r_0$ cannot be an outer black hole horizon \cite{GibHaw77}  (e.g., it is a cosmological one for pure dS spacetime, i.e., when $a_0^2=3/\Lambda>0$). For the Schwarzschild-AdS black hole the temperature is given by $T=\frac{1}{4\pi a_0}(1-\Lambda a_0^2)$ and the entropy by ${\cal S}=\pi a_0^2$ \cite{HawPag83}.

\subsection{Properties of the solution}

\label{subsec_prop}

\subsubsection{Geometry and photon sphere}

\begin{figure}[p]
	\centering
	\includegraphics[width=.6\textwidth]{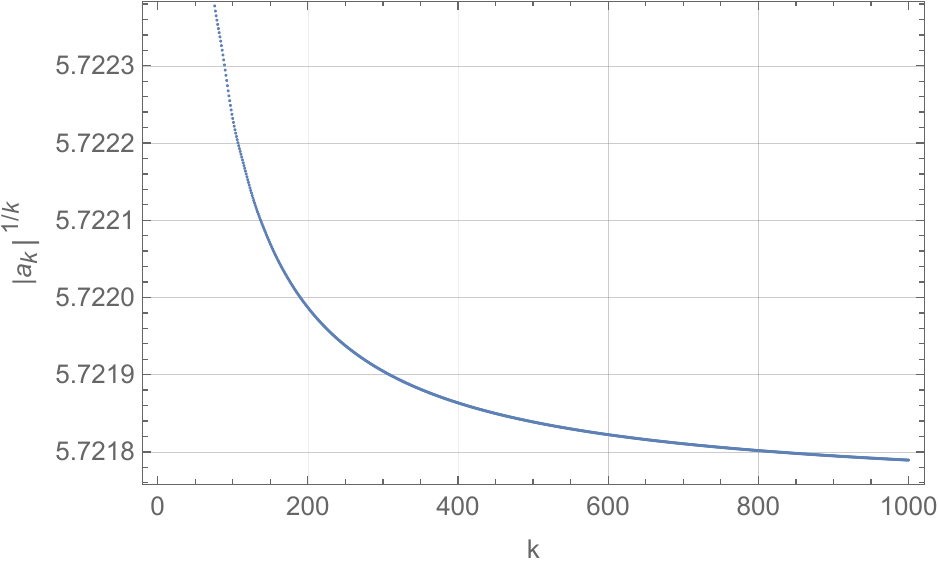}
	\caption{\small~Plot of $|a_k|^{\frac{1}{k}}$ against $k$ for $k=1,\ldots,1000$ (cf.~\eqref{a_k}) with parameters given by $\Lambda=-1/7$, $a_0=1$, $b_0=1/3$, $c_0=-1/5$ (such that $a_0^2-b_0^2>0$ and~\eqref{hor_conds} is satisfied). From the approximate asymptotic value of $|a_k|^{\frac{1}{k}}$ one can estimate the convergence radius of the power series for $\Omega$ (eq.~\eqref{expansions} with $n=0$) using the standard root test, which thus constraints the range of $\Delta$ (for the given choice of parameters) as $|\Delta|\lesssim 0.1748$.}
	\label{fig_root_a}
\end{figure}

\begin{figure}[p]
	\centering
	\includegraphics[width=.6\textwidth]{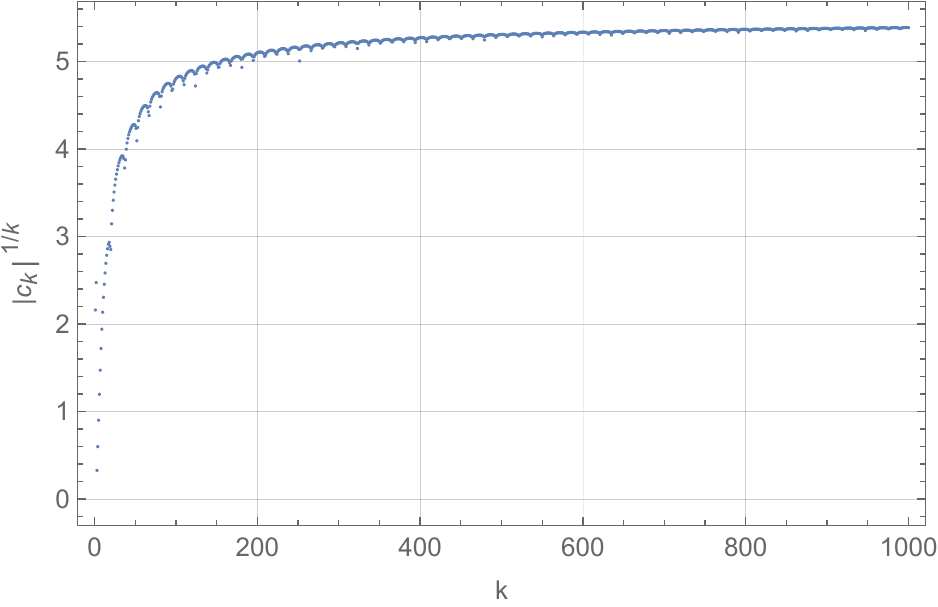}
	\caption{\small~The same as in figure~\ref{fig_root_a} (with the same choice of parameters) but now for the root $|c_k|^{\frac{1}{k}}$ (cf.~\eqref{c_k}). In this case, the root test indicates a slightly larger convergence radius for the power series for $H$ (eq.~\eqref{expansions} with $p=1$), namely $|\Delta|\lesssim 0.1856$. A similar estimate for the convergence radius of the power series for $\psi$ (eq.~\eqref{expansions} with $q=0$) can be obtained using the ratio $|b_k|^{\frac{1}{k}}$ (cf.~\eqref{b_k}) -- we omit the corresponding plot since it would not add relevant new information. For simplicity, in the rest of the paper we will always refer to the ``safer'' convergence radius $|\Delta|\lesssim 0.1748$ obtained in figure~\ref{fig_root_a}.}
	\label{fig_root_c}  
\end{figure}

The convergence properties of the series~\eqref{expansions} with \eqref{a1}--\eqref{b_k} can be analysed by standard methods. In figures~\ref{fig_root_a} and \ref{fig_root_c} we give evidence that the roots $|a_k|^{\frac{1}{k}}$ and $|c_k|^{\frac{1}{k}}$ tend to constant values for a large $k$ (the same is true for $|b_k|^{\frac{1}{k}}$, not displayed) for specific values of the three arbitrary parameters $a_0$, $b_0$ and $c_0$ and of $\Lambda$. Using the root test (cf, e.g., \cite{Rudinbook}) one can thus estimate the radius of convergence of each of the series~\eqref{expansions}. We have found a qualitatively similar behaviour also for some other values of the parameters.

\begin{figure}[p]
	\centering
	\includegraphics[width=.6\textwidth]{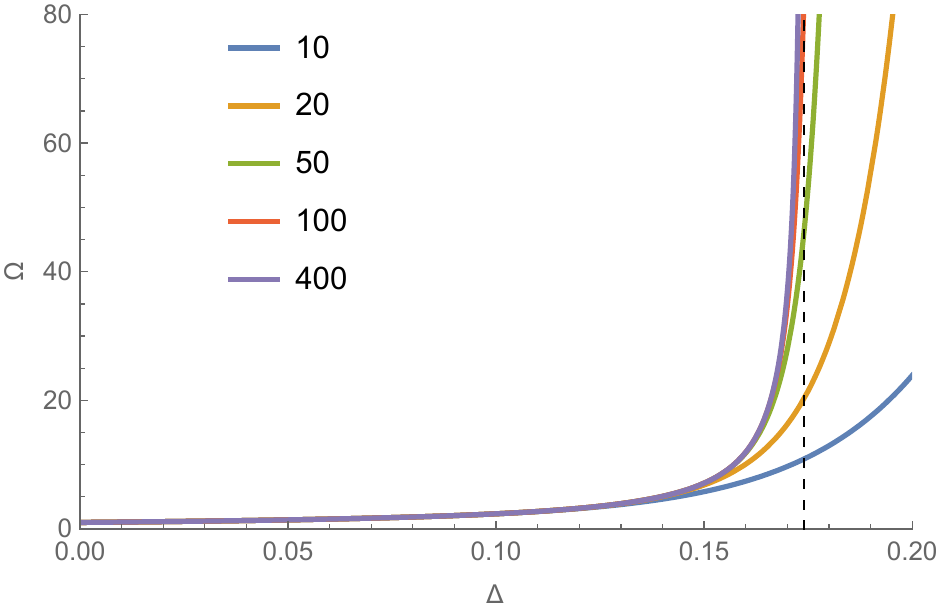}
	\caption{\small~Plot of $\Omega$ (eq.~\eqref{expansions} with $n=0$) against $\Delta\equiv r-r_0$ in the exterior region $\Delta>0$ with parameters as in figure~\ref{fig_root_a}. The plots are based on expansions taking into account the first 10, 20, 50, 100 and 400 terms, as indicated by the different colors. The event horizon is located at $\Delta=0$. The dashed vertical line represents the radius of convergence as estimated in figure~\ref{fig_root_a}. The fact that $\Omega$ grows very rapidly as one approaches the radius of convergence (and as more terms of the series are kept into account) suggests that the radius of convergence (in the exterior region) is close to conformal infinity. For example, keeping into account the first 100 terms in~\eqref{expansions} one obtains $\Omega\approx102$ near the convergence radius (and bigger values if more terms are summed in the series).}
	\label{fig_AdS3_Om} 
\end{figure}

\begin{figure}[p]
	\centering
	\includegraphics[width=.6\textwidth]{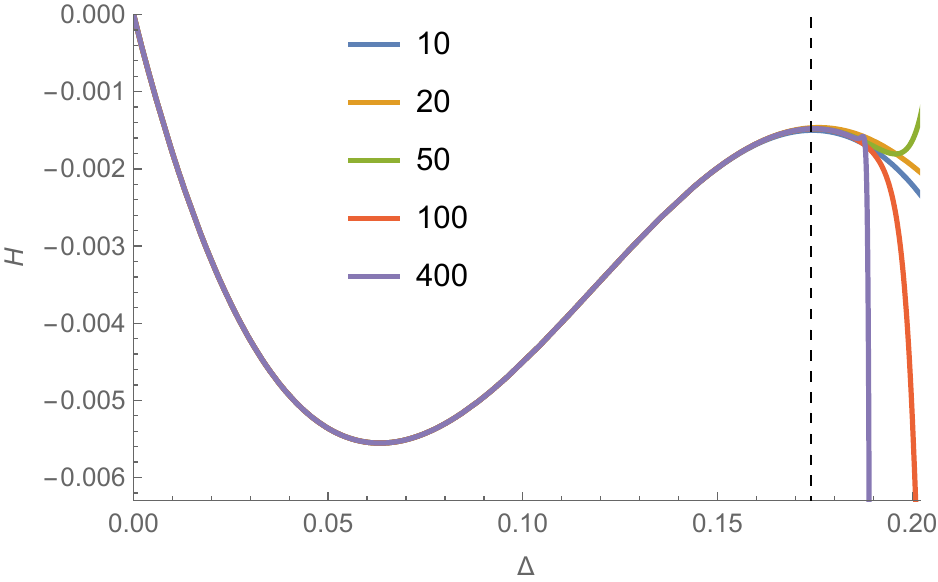}
	\caption{\small~The same as in figure~\ref{fig_AdS3_Om} but here for the plot of $H$ (eq.~\eqref{expansions} with $p=1$) in the exterior region. The local minimum of $H$ located at $\Delta\simeq0.063$ identifies a photon sphere \cite{ClaVirEll01,VirEll02} (cf. section~\ref{subsec_confK}).
		\label{fig_AdS3_H}} 
\end{figure}

The behaviour of the metric functions $\Omega$ and $H$ in the exterior region $\Delta>0$ and up to the convergence radius is depicted in figures~\ref{fig_AdS3_Om} and \ref{fig_AdS3_H}. The former indicates that (at least for certain choices of the integration constants) the convergence of the series breaks down near conformal infinity, where $\Omega$ blows up. The latter shows that $H$ is regular and enables one to identify a photon sphere \cite{ClaVirEll01,VirEll02} at the local minimum (see section~\ref{subsec_confK}). Furthermore, having determined the $r$-dependence of $H$, one can compute the angular radius of the black hole shadow~\eqref{shadow} for a static observer at any $r=r_O>a_0$ (within the convergence radius).

\begin{figure}[t]
	\centering
	\includegraphics[width=.8\textwidth]{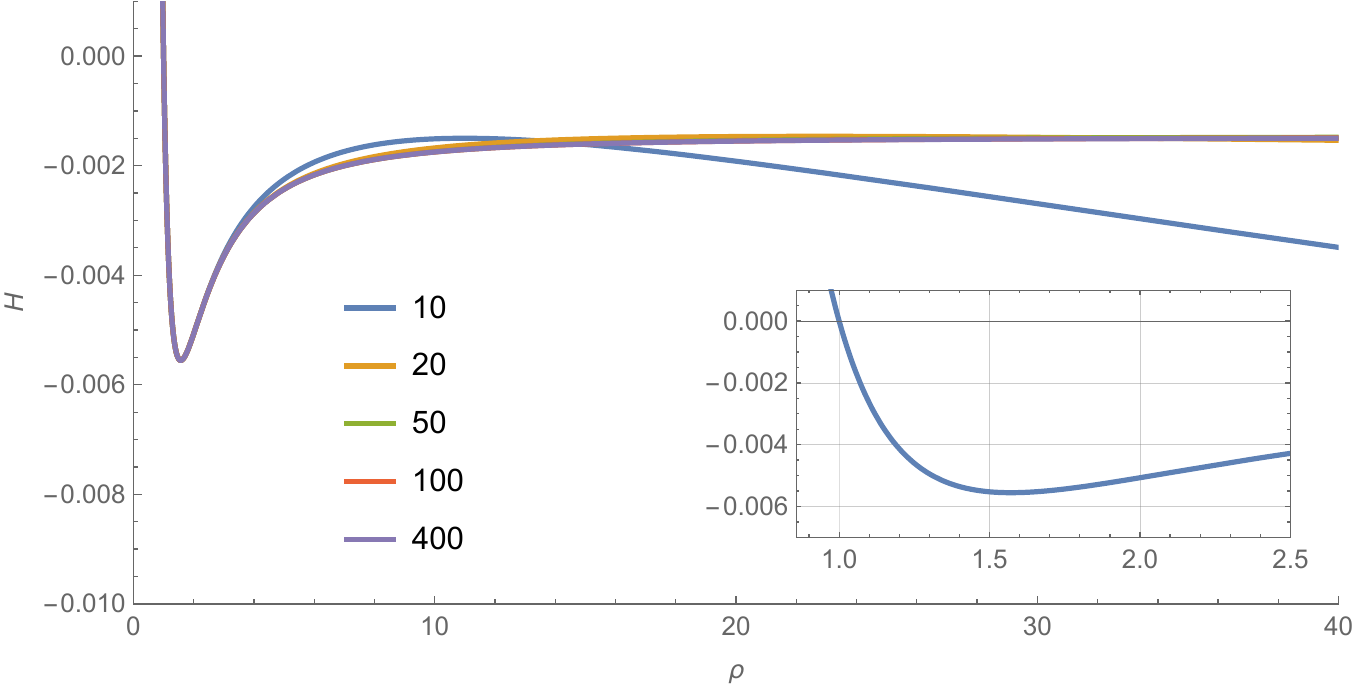}
	\caption{\small~The same as in figure~\ref{fig_AdS3_H} but $H$ is now plotted against the  Schwarzschild radial coordinate $\rho$, cf.~\eqref{metric_Schw}, \eqref{transf_Schw}.  The event horizon is located at $\rho=1$, where $H=0$, and the photon sphere (cf. the inset) at $\rho\simeq1.571$. The ``anomalous'', decreasing part of the curve based on an expansion with only the first 10 terms corresponds to a region well beyond convergence radius, cf. figure~\ref{fig_AdS3_Om}, and should thus not be taken into account. On the other hand, expansions with a larger number of terms indicate that $H$ tends asymptotically to a constant -- i.e., the metric function $f$ (eq.~\eqref{functions_Schw}) behaves as $f\sim\rho^2$ for a large $\rho$, as expected in an asymptotically AdS spacetime.
		By considering a series with up to 1000 terms (not displayed here) we have obtained a rough estimate of the asymptotic value of $H\approx-0.00148$ (this is, however, an extrapolation based on the value of $H$ at a distance from the horizon of about $10^3$ horizon radii -- a more accurate estimate may require a higher number of terms in the series, as well as a more precise knowledge of the radius of convergence).}
	\label{fig_AdS3_H(rho)} 
\end{figure}

Form a physical viewpoint, however, it is more interesting to characterize the dependence of $H$ as a function of the Schwarzschild radial coordinate $\rho$ (cf.~\eqref{metric_Schw}, \eqref{transf_Schw}) in the exterior region $\rho>a_0$, which is done in figure~\ref{fig_AdS3_H(rho)}. First, this enables one to identify the physical radius $\rho_{ps}$ of a photon sphere. While for vacuum black holes the ratio between $\rho_{ps}$ and the mass $M$ is exactly 3 (and thus independent of $\Lambda$, cf. \cite{StuHle99,ClaVirEll01} and  section~\ref{subsec_SchwAdS}, and \cite{Hilbert24} for the case $\Lambda=0$), our findings indicate that the effect of the scalar hair is to make $\rho_{ps}/M$ smaller, at least in the considered region of the parameter space (cf. table~\ref{tab_values}, where the choice $a_0=1$ means a unit horizon radius; the definition of $M$ is discussed below in the rest of this section). It can be further noticed that the general upper and lower bounds on $\rho_{ps}$ discussed in \cite{Hod13,CveGibPop16,LuLyu20,Hod20,YangLu20,Chakraborty21} are indeed fulfilled.\footnote{It should be emphasized, however, that references \cite{Hod13,CveGibPop16,LuLyu20,Hod20,YangLu20,Chakraborty21} typically assume the weak or even the dominant energy conditions, which can be violated in certain spacetime regions by a black hole sourced by a conformal scalar field (see appendix~\ref{app_energy} for more details). Nevertheless, since the same bounds as obtained \cite{Hod13,CveGibPop16,LuLyu20,Hod20,YangLu20,Chakraborty21} appear to be satisfied also here, it would be interesting if those proofs could be extended to cover also the case of a conformal scalar field -- however, this  goes beyond the scope of the present paper.} For comparison, let us observe that the photon sphere of both the BBMB and MTZ black holes is located at $\rho_{ps}=2M$ (cf.~\cite{ClaVirEll01,TomShiIzu17} for the BBMB case).\footnote{In the case $\Lambda=0$, photon spheres and shadows of black holes with a constant conformal scalar field \cite{DotGleMar08,NadVanZer08,AnaMae10,Astorino13} have been studied in \cite{Khodadietal20}. For more general asymptotically flat hairy black holes, a bound on the size of the ``hairosphere'' was obtained in \cite{NunQueSud96} and related to the size of the photon sphere in \cite{Hod11}, cf. also \cite{GhoSkSar23,IshMatIon24}. Some results for the case $\Lambda\neq0$ can be found in \cite{CaiJi98,IshMatIon24}.}	

\begin{figure}[p]
	\centering
	\includegraphics[width=.6\textwidth]{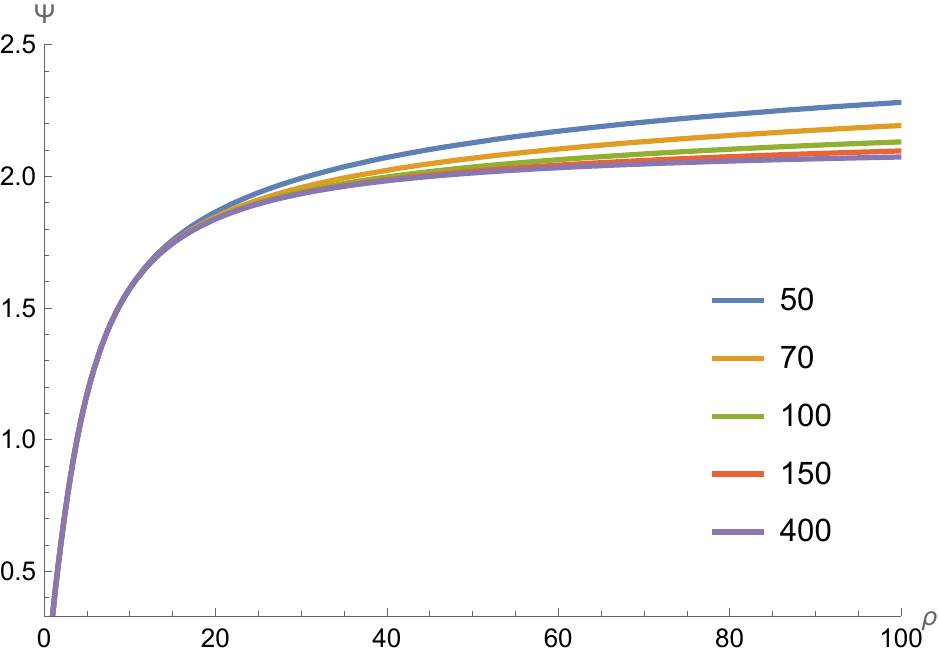}
	\caption{\small~Plot of the scalar field $\Psi$ (eqs.~\eqref{Psi}, \eqref{expansions}) as a function of the radius in Schwarzschild coordinates~\eqref{metric_Schw},\eqref{transf_Schw}  		
		 with parameters chosen as in figure~\ref{fig_root_a}. The plots are based on expansions taking into account the first 50, 70, 100, 150 and 400 terms, as indicated by the different colors. The event horizon is located at $\rho=1$, where $\Psi=b_0=1/3$. For the graphs with 100 and more terms, the displayed range of $\rho(=\Omega)$ lies fully within the convergence radius. The fact that $\Psi$ tends asymptotically to a constant means that the physical scalar fields goes for large $\rho$ as $\phi\sim 1/\rho$, in agreement with \cite{Winstanley03,RadWin05}. By considering a series with up to 1000 terms (not displayed here) we have obtained a rough estimate of the asymptotic value of $\Psi\approx2.13$ (this is, however, an extrapolation based on the value of $\Psi$ at a distance from the horizon of about $10^3$ horizon radii, cf. similar comments in figure~\ref{fig_AdS3_H(rho)}). It also follows from the graphs that $\Psi$ (and thus $\phi$) does not possess any nodes in the exterior region, again in agreement with the numerical results of \cite{Winstanley03,RadWin05}.}
	\label{fig_AdS3_Psi} 
\end{figure}

In addition, figure~\ref{fig_AdS3_H(rho)} shows the typical AdS behaviour $H\to$const$<0$ for $\rho\to\infty$. Using an expansion with a large number of terms one can obtain an estimate of the asymptotic value of $H$, and thus  compute the angular radius of the black hole shadow~\eqref{shadow} for a static observer close to infinity, cf. table~\ref{tab_values} (it will also be needed in section~\ref{subsubsec_thermod} to normalize the timelike Killing vector and thus compute the temperature). Similarly, the behaviour of the scalar field $\Psi$ in the exterior region is depicted in figure~\ref{fig_AdS3_Psi}, in qualitative agreement with the numerical findings of \cite{Winstanley03,RadWin05}. Also here one can estimate the asymptotic value of~$\Psi.$

\begin{table}[thbp]
	\[
	\begin{array}{|c|c||c|c|c|c|c|c|} \hline
		\  \Lambda \  & b_0 & \rho_{ps}\simeq & M\approx  & \rho_{ps}/M \approx & \sin^2\chi_{_O}^\infty\approx & T\approx & {\cal S}= \\ \hline\hline
		\multirow{5}{*}{$-\frac{1}{7}$} & 0 & 1.571 & 0.52 & 3.02 & 0.261 & 0.091 & \pi \\ 
		& 1/3    & 1.571 & 0.77 & 2.04 & 0.267 & 0.090 & 2.79  \\ 
		& 0.600  & 1.570 & 1.33 & 1.18 & 0.282 & 0.089 & 2.01  \\ 		
		& 0.857  & 1.566 & 2.16 & 0.73 & 0.324 & 0.087 & 0.83 \\ 
		& 0.970  & 1.578 & 2.69 & 0.59 & 0.388 & 0.099 & 0.19  \\ \hline
		
		\multirow{5}{*}{$-2$} & 0 & 2.500 & 0.83 & 3.01 & 0.926 & 0.239 & \pi \\ 
		& 1/3    & 2.468 & 0.87 & 2.84 & 0.924 & 0.245 & 2.79  \\
		& 0.600  & 2.386 & 0.95 & 2.51 & 0.922 & 0.263 & 2.01  \\ 			 
		& 0.857  & 2.236 & 1.14 & 1.96 & 0.917 & 0.336 & 0.83 \\ 
		& 0.970  & 2.200 & $\ding{53}$ & $\ding{53}$ & $\ding{53}$ & $\ding{53}$ & 0.19  \\ \hline
		      
		\multirow{5}{*}{$-3$} & 0 & 3.000 & 1.00 & 3.00  & 0.964 & 0.318 & \pi  \\ 
		& 1/3   & 2.941 & 1.01 & 2.91 & 0.963 & 0.327 & 2.79   \\ 
		& 0.600 & 2.795 & 1.05 & 2.66 & 0.960 & 0.356 & 2.01  \\ 			 		
		& 0.857 & 2.535 & 1.17 & 2.17 & 0.954 & 0.467 & 0.83 \\ 
		& 0.970 & 2.421 & $\ding{53}$ & $\ding{53}$ & $\ding{53}$ & $\ding{53}$ & 0.19  \\ \hline      		  
	\end{array}
	\]
	\caption{\small~The table shows the values attained by the physical quantities which characterize the AdS black hole solution~\eqref{a1}--\eqref{c_k} for different choices of $\Lambda$ and of the hair parameter $b_0$. In particular, the shadow angular radius (cf.~\eqref{shadow}) for a static observer close to infinity is given by $\chi_{_O}^\infty$. In all cases we have chosen $a_0=1$ and fixed the gauge as $c_0=-1/5$ (as in previous graphs) and we have approximated the solution by considering series~\eqref{expansions} with 1000 terms. The particular choice $\Lambda=-3$, $b_0=0.857$ corresponds to the case considered in figure~2 of \cite{RadWin05} (up to a different normalization of the scalar field and a different gauge choice in \cite{RadWin05}, which does not affect the displayed physical quantities). For comparison, vacuum Schwarzschild-AdS black holes, for which  $b_0=0$, are also included (the corresponding exact expressions, given in section~\ref{subsec_SchwAdS}, have been rounded up here to a limited number of decimal digits, which explains why the displayed ratio $\rho_{ps}/M$ does not equals the exact value~3 \cite{StuHle99,ClaVirEll01}). The symbol $\simeq$ means that the displayed values of $\rho_{ps}$ are approximate (only) because a finite number of terms are considered in the series~\eqref{expansions}, whereas $\approx$ is used for quantities that, in addition, involve an extrapolation to large values of $\rho$ (cf. the main text and figures~\ref{fig_AdS3_H(rho)} and \ref{fig_AdS3_Curur} for further comments).
		The values obtained for the latter should thus be taken with some caution and will need to be confirmed by different methods. By contrast, the displayed values of the entropy~${\cal S}$ (which does not depend on $\Lambda$) are based on the exact formula~\eqref{entropy} and thus do not involve any approximation (other than rounding it up to a limited number of decimal digits). In the last row of both cases $\Lambda=-2$ and $\Lambda=-3$, the symbols \ding{53} denotes quantities which we have been unable to compute using the series expansion method. The reason is that, for those particular values of the parameters, the convergence radius is reached well before $\rho$ can approach infinity -- the behaviour of $H$ and its derivatives (undisplayed) near the convergence radius further suggests that the corresponding spacetimes are not asymptotically simple \cite{penrosebook2}. This is presumably due to $b_0$ being too close to the critical value $b_0=1$, and seems compatible with the rapid growth of the mass observed in \cite{RadWin05} when $b_0\to1$.} 
	\label{tab_values}
\end{table}

\begin{figure}[t]
	\centering
	\includegraphics[width=.6\textwidth]{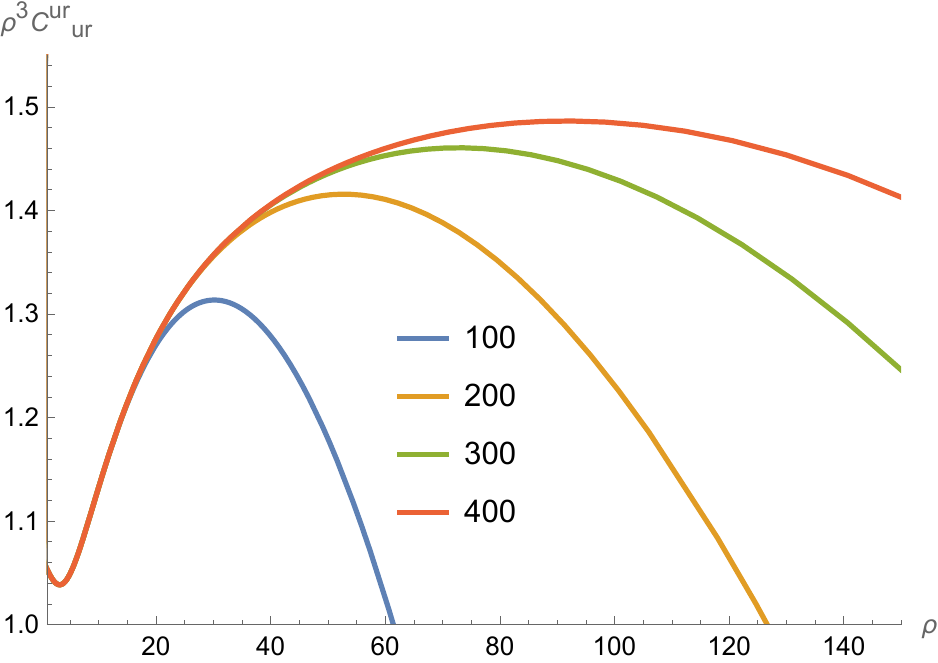}
	\caption{\small~For any asymptotically AdS metric of the form~\eqref{metric_Schw}, the Ashtekar-Magnon mass formula \cite{AshMag84} boils down to $M=\frac{1}{2}\rho^3C^{ur}_{\ \ ur}|_{\rho\to\infty}$ (cf., e.g., \cite{LuPopWen15}), where $C^{ur}_{\ \ ur}$ is a component of the Weyl tensor (which falls off asymptotically as $C^{ur}_{\ \ ur}\sim\rho^{-3}$). This is plotted as a function of $\rho$ using expansions with 100, 200, 300 and 400 terms, as indicated by the different colors, and the parameters are chosen as in figure~\ref{fig_root_a}.
		 The ``flat'' part of each curve (which in all cases falls within the convergence radius, cf. figures~\ref{fig_root_a} and \ref{fig_AdS3_Om}) indicates a spacetime region where our series correctly reproduces the asymptotic behaviour $\rho^3C^{ur}_{\ \ ur}\to$const. From a series with 1000 terms (not displayed) one can thus extrapolate, in a region around $\rho\approx 200$, an Ashtekar-Magnon mass given approximately by $M\approx 0.77$. Similarly, by choosing the parameters $\Lambda=-3$, $a_0=1$, $b_0=0.857$ (which correspond to those used in figure~2 of \cite{RadWin05}) one would estimate $M\approx 1.17$ (cf. table~\ref{tab_values}), which agrees within 2$\%$ with the numerical value obtained in \cite{RadWin05} (see also footnote~\ref{footn_mass}).}
	\label{fig_AdS3_Curur} 
\end{figure}

A few comments on the notion of mass used above and its values exemplified in table~\ref{tab_values} are now in order. In an asymptotically AdS spacetime, one can compute the mass $M$ using the Ashtekar-Magnon formula \cite{AshMag84} (provided the matter fields have a suitable fall-off \cite{AshMag84,AshDas00}; cf. \cite{AnaAstMar15} for related comments in the presence of a scalar field). However, making computations at conformal infinity with the series method illustrated above means going close or beyond the convergence radius of the solution and becomes therefore extremely challenging. As an example of these difficulties, we plot in figure~\ref{fig_AdS3_Curur} the quantity $\rho^3C^{ur}_{\ \ ur}$, which (for an asymptotically AdS metric of the form~\eqref{metric_Schw}) should tend to a constant equal to $2M$ for $\rho\to\infty$ \cite{AshMag84}. Figure~\ref{fig_AdS3_Curur} clearly shows that our expansions are unable to reproduce the correct asymptotic behaviour for an arbitrarily large $\rho$.\footnote{One arrives at a similar conclusion also by studying the behaviour of the dimensionless quantity $\frac{R}{4\Lambda}-1$, which departs from being zero (the value which should be attained by an exact solution, cf.~\eqref{Einstein}) as the value of $\rho$ grows ``too large'' (for example, for a series with 1000 terms, an error no larger than $1\%$ is obtained for $\rho\lesssim122$).} Nevertheless, for each of the curves there exists a range of $\rho$ (within the convergence radius, cf. figures~\ref{fig_root_a} and \ref{fig_AdS3_Om}) where the curve flattens, thus indicating a spacetime region where the series solution becomes close to reproducing the correct asymptotic behaviour. By considering an expansion with a large number of terms, one may thus attempt an extrapolation to compute (a lower bound on) the Ashtekar-Magnon mass -- cf. table~\ref{tab_values}. It follows that, at least for the considered values of the parameters, $M$ grows monotonically with $b_0$, in agreement with figure~5 of \cite{RadWin05}. We should stress, however, that further analysis with different (e.g. numerical) methods will be needed to set such kind of estimates on a firmer basis. This goes beyond the scope of this paper.\footnote{For values of the parameters similar to those used in figure~2 of \cite{RadWin05} (i.e., $\Lambda=-3$ and $b_0=0.857$, with our normalization), we extrapolated a value of the Ashtekar-Magnon mass in agreement within 2$\%$ with the numerical value obtained in \cite{RadWin05} (see also figure~5 therein), cf. the captions to our figure~\ref{fig_AdS3_Curur} and table~\ref{tab_values}. For the smaller values $b_0=0.6$ and $b_0=1/3$ of table~\ref{tab_values} (still with $\Lambda=-3$), we compared $M$ with further numerical data provided to us privately by Eugen Radu (see also figure~5 of \cite{RadWin05}) and found an agreement within 1$\%$. However, one would obviously need to test whether this agreement still holds on a larger set of values of the parameters.\label{footn_mass}}

\subsubsection{Thermodynamics}

\label{subsubsec_thermod}

Wald's formula enables one to compute the entropy in any diffeomorphism invariant Lagrangian theory \cite{Wald93}. For Lagrangians involving no more than second derivatives of the metric and first derivatives of the matter fields, one has \cite{Visser93,JacKanMye94,IyeWal94} 
\be
{\cal S}=-2\pi\oint Y^{abcd}\hat\epsilon_{ab}\hat\epsilon_{cd} , \qquad Y^{abcd}\equiv\frac{\partial {\cal L}}{\partial R_{abcd}} ,
\label{entropy_gen}
\ee
where the integral is taken with respect to the induced volume element on an arbitrary cross section of the horizon, and $\hat\epsilon_{ab}$ is the binormal to the cross section.  For
the theory~\eqref{I}, \eqref{S} (recalling footnote~\ref{footn_kinet}) one has $16\pi Y^{abcd}=(1-\phi^2)g^{a[c}g^{d]b}$, while a horizon at $r=r_0$ for metric~\eqref{metric} means $\mbox{\boldmath{$\hat\epsilon$}}=\Omega^2\d u\wedge\d r$. Then one readily finds
\be
{\cal S}=\frac{{\cal A}_h}{4}\left(1-\frac{b_0^2}{a_0^2}\right) ,
\label{entropy}
\ee
where ${\cal A}_h=4\pi a_0^2$ is the horizon area. The scalar hair thus results in a multiplicative factor which affects the standard area law, making the entropy smaller for a given horizon radius. As mentioned in section~\ref{subsec_010_general}, ${\cal S}$ is positive precisely in the branch $a_0^2-b_0^2>0$. Additionally, let us emphasize that \eqref{entropy} is a closed-form expression (since computed at the horizon, where only the first term of each of the series~\eqref{expansions} plays a role), from which one can obtain exact numerical values of the entropy for particular choices of the horizon radius $a_0$ and the hair parameter $b_0$. For example, choosing the parameters as in figure~\ref{fig_root_a} one obtains ${\cal S}=\frac{8\pi}{9}$  (cf. table~\ref{tab_values}). Modifications to the area law due to non-minimally coupled scalars have been pointed out previously in \cite{CreMan95,AshCorSud03} (see also \cite{BarDohWin05,MarTroSta06} for specific examples, and \cite{MarZan96,Henneauxetal02} for earlier results in $2+1$ dimensions).\footnote{It may be useful to further observe that formula~\eqref{entropy} has been obtained off-shell, and is not modified by adding a self-interaction potential to the theory~\eqref{I}, cf.~\cite{NadVanZer08}. Indeed, when specialized to the particular case of the $\Lambda>0$ MTZ black hole \cite{MarTroZan03} (for which $a_0^2-b_0^2<0$ \cite{DotGleMar08}), it produces a result in agreement with \cite{BarDohWin05}, including the negative black hole horizon entropies discussed there (see~\cite{NadVanZer08} for further comments).
}

Finally, let us compute the temperature associated with a generic Killing vector field $\xib=N\pa_t=-N\pa_u$, for the time being without specifying the normalization constant $N>0$ ($\xib$ has norm $\xi_b\xi^b=N^2\Omega^2H$ and is thus timelike future oriented where $H<0$, and null at the horizon, i.e., for $H=0$). The surface gravity is defined by $\nabla^a(\xi_b\xi^b)\doteq-2\kappa\xi^a$ \cite{Waldbook}, where the symbol $\doteq$ denotes equality at the horizon. For the temperature $T=\frac{\kappa}{2\pi}$ one thus finds 
\be
T=-\frac{N c_0}{4\pi} .
\label{T_gen}
\ee
Next, we normalize the Killing field such that $\xi_b\xi^b\sim\frac{\Lambda}{3}\rho^2$ for $\rho\to\infty$ \cite{HawPag83} (cf. \cite{BroCreMan94} for further comments). For a specific choice of parameters of the solution, one can employ the asymptotic (approximate) value of $H$ to fix $N$ (figure~\ref{fig_AdS3_H(rho)}) and thus estimate a numerical value of $T$. For example, with the parameters of figure~\ref{fig_root_a} one obtains $T\approx0.091$ (cf. table~\ref{tab_values} for different choices of parameters).

\section{Conclusions}

\label{sec_concl}

 We have studied analytically static, spherically symmetric solutions to AdS-Einstein gravity conformally coupled to a scalar field, in the absence of any self-interaction potential. In the first part of our contribution, we have described how to reduce the field equations to the simple form~\eqref{uu_2}--\eqref{scalar4} by a choice of suitable coordinates (cf. also \cite{Ray24}, and \cite{PraPraPodSva17,Podolskyetal18} in a different context). This first result, employed in the rest of the paper, could be useful also in various future studies. Next, we have set up an ansatz~\eqref{expansions} for a power series solution of the equations~\eqref{uu_2}--\eqref{scalar4}. By analyzing the corresponding indicial equations, we have identified distinct branches of solutions, summarized in table~\ref{tab_exponents}. These correspond to expansions at physically different  points and may admit different numbers of integration constants, thus describing physically distinct solutions. While a detailed study of all possible cases will deserve a separate investigation, in the rest of the paper we have focused on AdS black holes, by considering an expansion around a (necessarily non-extremal) Killing horizon. This gave rise to a solution admitting two integration constants, corresponding to the horizon radius and the value of the scalar field $\Psi$ at the horizon. When the scalar hair vanishes, one recovers the Schwarzschild-AdS vacuum black hole. Properties of the solutions have been described, also using various plots, and compared with the numerical findings of \cite{Winstanley03,RadWin05}. We have, in particular, shown how the coordinates employed throughout this work permit naturally to localize a photon sphere and compute its radius, which is affected by the the scalar field but fulfills known general bounds  \cite{Hod13,CveGibPop16,LuLyu20,Hod20,YangLu20,Chakraborty21}, as well as the associated shadow. In addition, we have briefly discussed the thermodynamics of the solutions. While the entropy can be computed exactly at the horizon, the obtained values of the mass and temperature rely on an extrapolation at large radii, and would need to be further studied by other methods (at least for certain values of the parameters, however, we have found good agreement with the numerics of \cite{RadWin05}).
 
 The methods used in this paper can be extended to other contexts, such as  black holes with non-spherical horizons, as well as certain extensions of the theory~\eqref{I}. It will be also interesting to study analytically the solitons found numerically in \cite{RadWin05}. This will be discussed elsewhere.

\section*{Acknowledgments}

I am grateful to Sourya Ray for collaboration at an early stage of this work, for helpful suggestions and for making available to me his work \cite{Ray24} before publication, and to Eugen Radu for useful comments about \cite{RadWin05} and for kindly sharing with me large samples of related numerical data. I also thank Vojt{\v e}ch Pravda for helpful discussions, and Tereza Lehe\v ckov\'a for pointing out a typo in eq.~\eqref{Zz} in the first draft of the manuscript. This work has been supported  by the Institute of Mathematics, Czech Academy of Sciences (RVO 67985840).

\renewcommand{\thesection}{\Alph{section}}
\setcounter{section}{0}

\renewcommand{\theequation}{{\thesection}\arabic{equation}}

\section{Energy-momentum tensor}
\setcounter{equation}{0}

\label{app_energy}

The energy-momentum tensor defined by the theory~\eqref{I} is given by (cf. \eqref{Einstein}) $T_{ab}=\phi^2\tilde G_{ab}$, where $\tilde G_{ab}$ is the Einstein tensor of the auxiliary metric $\tilde g_{ab}=\phi^2g_{ab}$. The spacetimes considered in this paper are of the form~\eqref{metric} with \eqref{Psi}, so that the line-element associated with $\tilde g_{ab}$ reads
\be
\d\tilde s^2=\Psi^2\left(-2\d u\d r+H\d u^2+\d\Sigma^2\right) , \qquad \d\Sigma^2=2P^{-2}\d\zeta\d\bar\zeta , \qquad P=1+\frac{1}{2}\zeta\bar\zeta .
\ee

After computing $\tilde G_{ab}$, the non-zero mixed components of $T^a_{\phantom{a}b}$ in the Schwarzschild coordinates~\eqref{metric_Schw}--\eqref{functions_Schw} read
\beq
& & \Omega^4 T^t_{\phantom{t}t}=-\left[\Psi^2+H'\Psi\Psi'+H(2\Psi''\Psi-\Psi'^2)\right] , \label{Ttt} \\
& & \Omega^4 T^\rho_{\phantom{\rho}\rho}=-(\Psi^2+H'\Psi\Psi'+3H\Psi'^2) , \\
& & \Omega^4 T^\zeta_{\phantom{\zeta}\zeta}=\Omega^4 T^{\bar\zeta}_{\phantom{\bar\zeta}\bar\zeta}=\Psi^2+\Psi(H\Psi')'+H\Psi'^2 , \label{TZz}
\eeq
where we have simplified~\eqref{TZz} using \eqref{scalar4}, such that $T^a_{\phantom{a}a}=0$ (see the comment at the end of section~\ref{sec_intro}).

The energy density is given by $-T^t_{\phantom{t}t}$ (cf., e.g., \cite{MayBek96}). By definition \cite{HawEll73}, the  weak energy condition (WEC) is satisfied in regions where $-T^t_{\phantom{t}t}\ge0$ along with $-T^t_{\phantom{t}t}+T^\rho_{\phantom{\rho}\rho}\ge0$ and   $-T^t_{\phantom{t}t}+T^\zeta_{\phantom{\zeta}\zeta}\ge0$. The dominant energy condition (DEC) requires that, in addition, also $-T^t_{\phantom{t}t}-T^\rho_{\phantom{\rho}\rho}\ge0$ and   $-T^t_{\phantom{t}t}-T^\zeta_{\phantom{\zeta}\zeta}\ge0$.   

\begin{figure}[t]
	\centering
	\includegraphics[width=.6\textwidth]{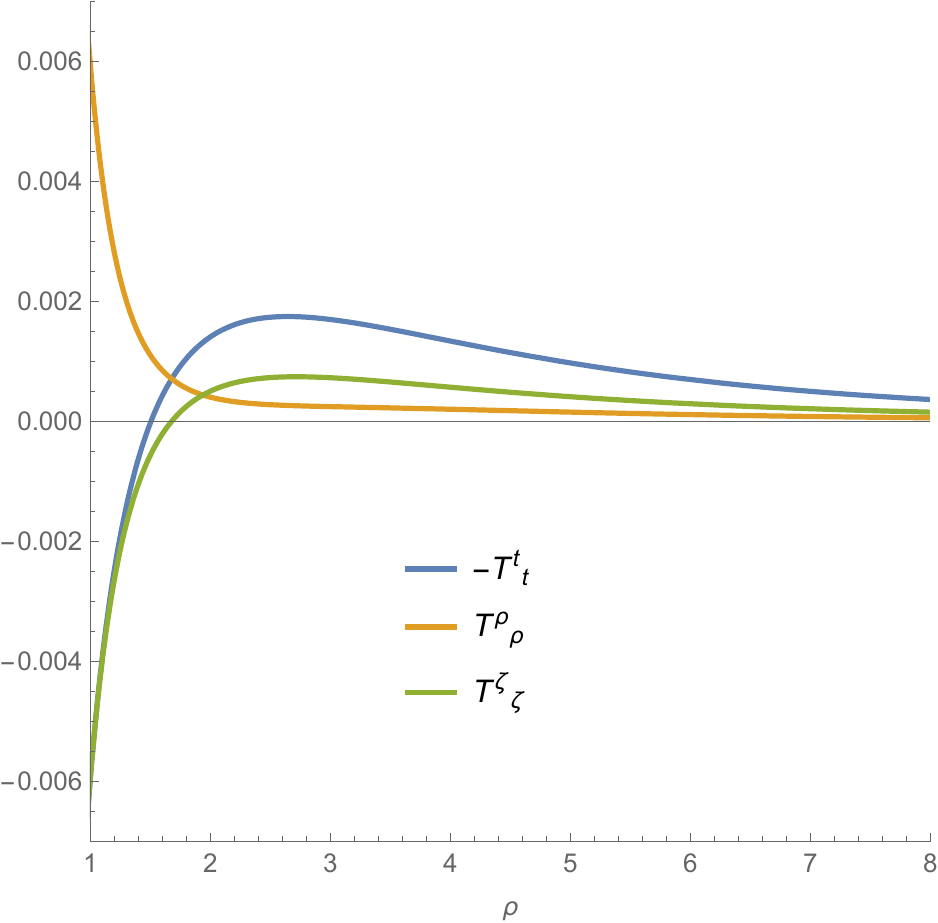}
	\caption{\small~Plot of the energy-momentum components~\eqref{Ttt}--\eqref{TZz} in the exterior region $\rho>1$ of the black hole solution~\eqref{a1}--\eqref{b_k} (with parameters as in figure~\ref{fig_root_a}). The plot is based on an expansion with 400 terms. From the graphs it follows that the energy density is positive for $\rho\gtrsim 1.506$, the WEC holds for $\rho\gtrsim 1.571$, while the DEC for $\rho\gtrsim 1.678$. A similar plot has been obtained numerically in figure~8 of \cite{Winstanley03}.}
	\label{fig_energy} 
\end{figure}

For the black hole solution of section~\ref{subsec_010_general} one finds at the horizon (i.e., for $H=0$; as in section~\ref{subsubsec_thermod}, the symbol $\doteq$ denotes equality at the horizon)
\be
  T^t_{\phantom{t}t}\doteq -\frac{\Lambda}{3}\frac{b_0^2}{a_0^2-b_0^2}>0 .
\ee 
Recall that for the hairy solution considered throughout the paper one has $a_0^2-b_0^2>0$, therefore the energy density is negative and thus the WEC is violated at (and, by continuity, in the vicinity of) the horizon -- in agreement with an observation of \cite{Winstanley03}. One further has $-T^\zeta_{\phantom{\zeta}\zeta}\doteq T^\rho_{\phantom{\rho}\rho}\doteq T^t_{\phantom{t}t}$ (cf. also \cite{NunQueSud96,MayBek96}). However, the plot in figure~\ref{fig_energy} reveals that (at least in a certain region of the parameter space) both the WEC and the DEC are satisfied sufficiently far from the horizon. In particular, the WEC holds precisely outside the photon sphere (at least within the accuracy of the series solution), cf. figure~\ref{fig_AdS3_H(rho)}.

\providecommand{\href}[2]{#2}\begingroup\raggedright\endgroup

\end{document}